\documentclass[useAMS,usenatbib]{mn2e}
\input epsf.sty
\usepackage{graphicx}

\def\feiiopt{{Fe \sc{ii}}$_{\rm opt}$\/}
\def\feii{{Fe\sc{ii}}$_{\rm opt}$\/}

\def\oiiiopt{{\sc{[Oiii]}}$\lambda\lambda$4959,5007\/}
\def\rfe{R$_{\rm FeII}$}
\def\feiiq{\rm Fe{\sc ii}$\lambda$4570\/}
\def\hb{{\sc{H}}$\beta$\/}
\def\hbbc{{\sc{H}}$\beta_{\rm BC}$\/}
\def\hbnc{{\sc{H}}$\beta_{\rm NC}$\/}

\def\lm{$\rm L/M$}
\def\Ms{M$_\odot$ }
\def\kms{km$\;$s$^{-1}$}

\def\civ{{\sc{Civ}}$\lambda$1549\/}
\def\mv{m$_{\rm V}$}
\def\mb{$\rm M_{\rm B}$}
\def\Mbh{M}
\def\PopA{Pop. A}
\def\PopB{Pop. B}
\def\oiii{{\sc [Oiii]}$\lambda$5007}
\begin{document}
 \title{Searching For the Physical Drivers of Eigenvector 1: \\
 Influence of  Black Hole Mass and Eddington Ratio}
 \author[Marziani, Zamanov,   Sulentic, \& Calvani]
   { Paola Marziani$^1$\thanks{E-mail:
    marziani@pd.astro.it (PM); zamanov@pd.astro.it (RZ)
    calvani@pd.astro.it (MC), giacomo@merlot.astr.ua.edu (JWS)}, Radoslav K. Zamanov$^{1}$\thanks{Present address:
    Astrophysics Research Institute, Liverpool John Moores University, Twelve Quays House, Egerton Wharf,  Birkenhead CH41 1LD
  United Kingdom; email: rz@astro.livjm.ac.uk},
      Jack W. Sulentic$^{2}$,  Massimo Calvani$^{1}$ \\
$^{1}$Istituto Nazionale di Astrofisica, Osservatorio Astronomico di Padova,
       Vicolo dell'Osservatorio 5, I-35122 Padova, Italy \\
$^{2}$ Department of Physics and Astronomy, University of
       Alabama, Tuscaloosa, AL 35487, USA
}


 %
\date{Received ...........; accepted ........} \maketitle
\begin{abstract}
We compute  the virial mass of the central black hole (\Mbh) and the
luminosity-to-mass (L/M) ratio of $\approx 300$ low-$z$\ quasars and luminous
Seyfert 1 nuclei. We analyze: (1) whether radio-quiet and radio-loud objects
show systematic differences in terms of M and L/M; (2) the influence of M and
L/M on the shape of the \hb\ broad component line profile; (3) the
significance of the so-called ``blue outliers" i.e., sources showing a
significant blueshift of the \oiiiopt\ lines with respect to the narrow
component of \hb\ which is used as an estimator of the quasar reference frame.
We show that \Mbh\  and L/M distributions for RQ and RL sources are likely
different for  samples matched in luminosity and redshift, in the sense that
radio-quiet sources have systematically smaller masses and larger L/M. However,
the L/M ratio distributions become indistinguishable if 8.5$<\log$M$<9.5$. Line
profile comparisons for median spectra computed over narrow ranges of M and
L/M indicate that a Lorentz function provides a better fit for higher L/M
sources and a double Gaussian for lower L/M values. A second (redshifted)
Gaussian component at low L/M appears as a red asymmetry frequently observed
in radio-loud  and radio-quiet sources with broader (FWHM$\ga$4000 \kms) \hb\
broad component profiles. This component becomes stronger in larger mass and
lower L/M sources. No specific influence of radio loudness on the \hb\ broad
component profile is detected, although  equivalent widths of \hb\ broad
component and especially of \oiiiopt\ are larger for radio-loud sources. We
identify five more  ``blue outlier" sources. Since these sources are, on
average, one magnitude brighter than other AGNs with similar mass, they are
accreting at an Eddington ratio that is 2-3 times higher.  We hint at
evolutionary effects that explain some of these results and reinforce the
``Eigenvector 1" correlations.
\end{abstract}




\section{Introduction}
Studies of emission lines play an important role in our
understanding of Active Galactic Nuclei (AGN) and the physics the
accretion processes, although a connection between the main
theoretical parameters  (i.e., black hole mass, accretion rates,
black hole spin, and a viewing angle) and observed spectral
parameters is, at best, just sketchy and qualitative. In recent
years, a major innovation has been ascribed to the so-called
``Eigenvector 1" (E1) correlations (Boroson \& Green 1992). The
original E1 was most closely related to the anti-correlation
between \feiiopt\ strength and peak \oiii\ intensity, and FWHM of
the \hb\ broad component (\hbbc).

E1-related correlations have emerged as  very robust, since they
continue to appear in different low-$z$ \ AGN samples, even if a
set of spectral parameters partly different from the ones used
for the original Principal Component Analysis (PCA)  by Boroson
\& Green (1992) is considered. An outstanding relationship  in
the context of E1 involves the FWHM(\hbbc) and the ratio between
the equivalent width of the \feii\ complex centered at
$\lambda$4570 and \hbbc, \rfe = W(\feiiq)/W(\hbbc). Low redshift
AGN occupy an elbow sequence in the plane defined by these two
parameters (see Sulentic et al. 2000a; Marziani et al. 2001). The
overall sequence is defined by RQ sources while the RL AGN occupy
only a restricted part of it. The domain where FWHM(\hbbc)$\la$
4000 \kms\ is predominantly occupied by RQ sources (Sulentic et
al. 2000a; see also Sulentic et al. 2003). There is an apparent
dichotomy in the \hbbc\ line profile shapes: sources with
FWHM(\hbbc) $\la$ 4000 \kms\ show typically a Lorentzian \hbbc,
while for FWHM(\hbbc) $\ga$ 4000 \kms\ a double Gaussian fit is
more appropriate (Sulentic et al. 2002); (3) the so-called ``blue
outliers" (sources showing a significant blueshift of the
\oiiiopt\ lines with respect to inferred quasar
 frame) occur among sources with strongest \feiiq\  narrowest
 \hbbc\ (Zamanov et al. 2002; Marziani et
al. 2003a).

The robustness of the E1 correlations stems probably from a series of
concurring factors: (1) an Eddington ratio that may be continuously changing
along the E1 sequence (e.g., Boroson, Persson \& Oke 1985; Boroson \& Green
1992; Murray et al. 1995; Nicastro 2000; Marziani et al. 2001; Boroson 2002);
(2) outflow/wind properties which are expected to depend on Eddington ratio
(e.g., Dultzin-Hacyan et al. 2000; Marziani et al. 2001); (3) orientation
effects that will blur any physical correlation and displace sources in the
direction of increasing Eddington ratio (Marziani et al. 2001; see also, Nagao
et al. 2000, Zhou et al. 2003); (4) Broad Line Region (BLR) structural effects
(Sulentic et al. 2000b, Sulentic et al. 2002; see also the Proceeding of the
Nebraska meeting for the E1 relevance for the Broad Line Region structure;
Gaskell et al. 1999); (5) last, but not least, evolutionary effects that will
lead to an increase of black hole mass as well as of the \oiii\ rest-frame
equivalent width (Zamanov et al. 2002; Boroson 2002). We will briefly discuss
them in \S~ \ref{nbo} and \S~9.

Even if there is not yet a consensus (see e.g., Marziani et al.
2001; Boroson 2002), the physical drivers of E1 are likely to
involve the luminosity-to-black hole mass ratio (L/M $\propto$
Eddington ratio), an orientation angle, and the black hole mass
(M) (Zamanov \& Marziani 2002).  Additional properties such as
black hole spin (e.g., Wilson \& Colbert 1995) or the properties
of the host galaxy environment (e.g., McLure et al. 1999) are also
likely to play a role in any RQ-RL  dichotomy.


This paper presents a basic investigation of the influence of L/M and M  on RL
and RQ sample properties, on the \hbbc\ line profile, and on the occurrence of
the ``blue outliers", which are all major E1-dependent phenomena. In \S~2 we
present our quasar sample followed by (\S~3) derivation of M and L/M, (\S~4)
presentation and discussion of mass -- luminosity diagrams, (\S~5) a
comparison of radio quiet and loud sources, (\S~6) an analysis of \hbbc\
profiles, (\S~7) as well as  significance of the ``blue outlier" sources. Our
results are discussed in \S~8. \S~9 provides some concluding remarks.

\begin{figure}
\mbox{} \vspace{9.0cm} \includegraphics{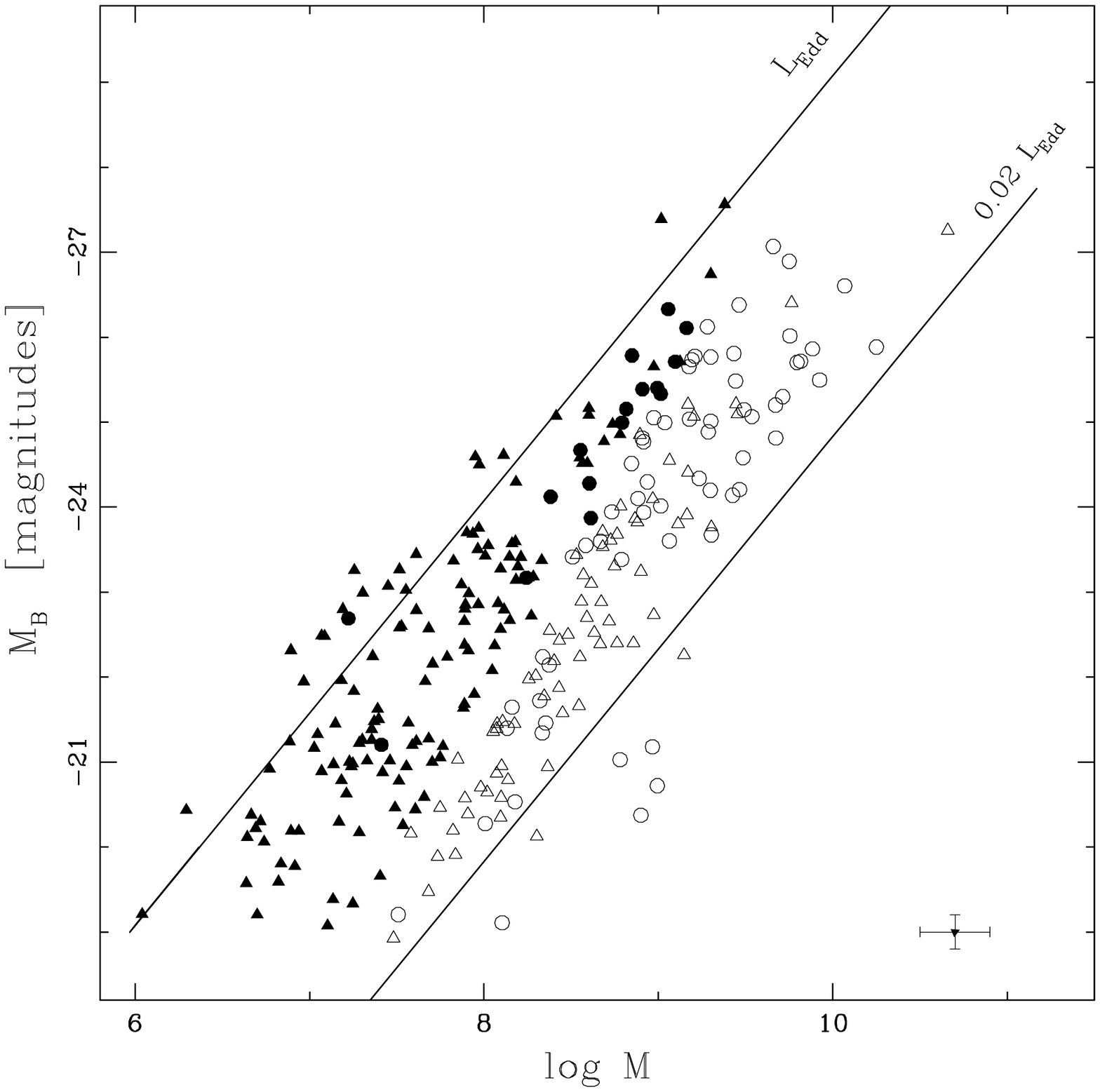} \caption[]{Distribution of absolute B magnitude (\mb)
versus black hole mass \Mbh\ (in solar units) for the 278 AGN sample. \mb\
values are corrected for the galactic extinction. Triangles and circles
indicate radio-quiet and radio-loud sources respectively. Open symbols
indicate sources with FWHM(\hbbc)$\ga$4000 \kms; filled symbols sources with
FWHM(\hbbc)$\la$4000 \kms (see \S \ref{popab}). The  bars in the lower right
corner indicate typical errors.} \label{sample}
\end{figure}

\section{Sample Definition \& Measurements}

We use spectra of 278 AGNs. Our  AGN sample includes 215 sources and has been
presented elsewhere (Marziani et al. 2003b, hereafter M03). We added 63 soft
X-ray selected sources with suitable optical spectroscopic data (Grupe et al.
1999; hereafter G99) because this turns out to be the most fertile source of
``blue outliers". Inclusion of Grupe et al. data makes our sample less biased,
as described in \S \ref{popab}.  The optical characteristics of these X-ray
selected sources include: (1) relatively ``narrow" broad emission lines (e.g.
Narrow Line Seyfert 1 [NLSy1] sources), (2) strong optical \feiiopt\ blends
and (3) weak forbidden emission lines. The merged datasets provide reasonable
quality spectra (S/N$\approx 20-70$) for  the region near \hb\ in 278 AGN ($z
\la $ 0.8). The M03 dataset is not complete (see M03 for discussion of
completeness and sample biases).

We use line parameters reported in M03 and derive the
corresponding measures (using the same procedures  described in
M03) for the X-ray selected sample from the  digital spectra made
publicly available by G99. We  consider here measurements of the
following parameters: (a) FWHM(\hbbc); (b)  equivalent width
ratio \rfe=W(\feiiq)/W(\hbbc), (c)  FWHM measurement of the lines
in the \feiiq\ blend (see M03 for details), (d)  radial velocity
difference between \oiii\ and \hb\ at the line peak (without
narrow component subtraction). Accuracy of the G99 wavelength
calibration  was evaluated by examining the distribution of
radial velocity difference between [OIII]$\lambda$4959 and
[OIII]$\lambda$5007. Whenever possible, [OIII]$\lambda$4363 was
also used  and all inter-narrow line differences were found to be
comparable to our measures (cf. Zamanov et al. 2002).

Apparent (\mv), and absolute B magnitudes (\mb) were taken from 10$^{\rm th}$
edition of the AGN catalog (V\'eron-Cetty \& V\'eron 2001). A magnitude
correction  $+0.88$ mag was applied to convert \mb\ from H$_0$=50
\kms~Mpc$^{-1}$ to H$_0$=75 \kms~Mpc$^{-1}$  (q$_0=0$) which we use throughout
this paper. Galactic extinction values $\rm A_B$\ were taken from NED
following Schlegel et al. (1998). Sources were considered RL if the specific
flux at 6 cm and in the B band was larger than 10, with data taken from the
V\'eron-Cetty \& V\'eron (2001) catalog.

%
%

\section{Computation of \Mbh\ and L/M}
\subsection{Black Hole Masses}

One can estimate the mass of the putative supermassive black hole using
FWHM(\hbbc) and a reverberation ``radius" (Kaspi et al. 2000) along with the
assumption of virialized motions. It is now common to estimate the mass by
assuming the BLR distance from the central continuum source $\rm r_{\rm
BLR}\propto (L_{5100})^{\alpha}$ and $\alpha=$0.7, as  derived from the
reverberation data (Kaspi et al. 2000). The virial mass is M=$\rm r_{\rm
BLR} v^2 / G$, where v= $\sqrt{3}/2 $ FWHM(\hbbc) (c.f. Woo \& Urry 2002a),
and G is the gravitational constant. We therefore can write the black hole
virial mass as follows:
\begin{equation}
  {\rm M} = 4.817 \times \rm  \left( \frac { FWHM(H\beta_{\rm BC})}
  {1\; km\:s^{-1}} \right) ^2
  \left( \frac {\lambda \: L_{5100}} {10^{44}\:erg\:s^{-1}}
  \right)^{0.7},
  \label{eqMass}
\end{equation}
where \Mbh\ is  in solar units, L$_{5100}$\ is the specific luminosity at
5100 \AA\ (in units of ergs s$^{-1}$ cm$^{-2}$ \AA$^{-1}$), and
 \begin{equation}
\rm    \lambda \: L_{5100} =
 3.137\times 10^{35-0.4({\rm M_B} - A_B )} \:erg\:s^{-1}.
 \end{equation}
In order to convert the absolute B magnitude \mb\  to $\lambda
L_{5100}$\ we assumed that the specific flux is $\rm f_\nu
\propto \nu^{-0.3}$.  The derived masses are thought to be
reasonable estimates within a factor of about 2--3 (see also Woo
\& Urry 2002a, Vestergaard 2002).

\subsection{Influence of Orientation}

Source viewing angle is especially important in AGN where the
concepts of disk accretion/line emission and jet ejection are
widely favored. In the above mass derivation formula we did not
include any orientation related broadening factor. If \hbbc\ is
emitted in a highly flattened structure then \Mbh\ can be
underestimated by a factor $\approx \sin i^{-2}$ in
pole-on/face-on sources (i.e., with accretion disk axis oriented
along our line of sight). Unfortunately, we do not yet have a
proper way to estimate the viewing angle $i$\ for each source.
This is a major drawback because there is observational evidence
suggesting that orientation can affect line width by a factor
$\approx$ 2 (Brotherton, 1996; Marziani et al. 2001; Jarvis \&
McLure, 2002). Even though we neglected the effects of
orientation in the general sample population, we considered it in
computations of \Mbh\ and L/M the  relatively rare and extreme
class that is likely observed close to ``pole-on":   the ``blue
outliers" (\S \ref{bo}).

\subsection{The Role of the BLR Size-Luminosity Relationship}

The exponent in the relationship between source luminosity L and $\rm
r_{\rm BLR}$ is assumed to be $\alpha$=0.7. Any deviation from this value
has quantitative effects on our results. Our sample includes many moderately
luminous quasars at $0.4 < z <0.8$, so it is important to stress  that the
relationship derived by Kaspi et al. (2000) is based exclusively on quasars
of $z< 0.4$ and that the high (and low) luminosity ranges of the
correlation are poorly sampled.

If we  consider  sources in the luminosity range $43.4 \la \log \rm L/L_\odot
\la 45$\  (i.e. where we have uniform luminosity sampling), the slope of the
best fit is $\alpha$=0.8, and could be easily as high as $\alpha$ = 1 without
increasing significantly the fit standard deviation. Marziani et al. (2001)
considered this case which seems appropriate  for the  PG quasar luminosity
range.  One must remain open to the possibility that $0.5 \la \alpha \la 1$,
and that $\alpha$\ might even be  a function of L. Changing $\alpha$\ implies
an L-dependent change in mass estimates; therefore the slope of the
luminosity-to-mass relationship (Fig. \ref{sample}) is affected as well as the
location of points in the L/M vs. M diagram (Fig. \ref{bins} \& Fig.
\ref{fig10}). Despite these possibilities, systematic trends discussed in this
paper should not be  affected.


\subsection{Bolometric Luminosity}

We  calculated the bolometric luminosity L from $\rm L \approx 10
\lambda\; \rm L_{\lambda} (5100 $\AA) (details can be found in
Wandel, Peterson, \& Malkan, 1999, Elvis et al. 1994, Collin et
al. 2002). Values computed in this way were used to derive the
\lm\ ratio. The L/M ratio everywhere is expressed in solar units
with the solar value (L/M)$_\odot$ = 1.92 ergs s$^{-1}\:$
g$^{-1}$, and the Eddington limit corresponding to $\log$L/M =
4.53.

We neglected any possible differences in the spectral energy
distribution between source subclasses. In order to test this
benign neglect we considered 44 sources that are common with Woo
and Urry (2002a) where SEDs were constructed from archival data.
Comparison of our bolometric luminosity values showed no
significant systematic difference: $ \overline{\Delta \rm \log L}
\approx -0.06$ for all  44 sources, with a standard deviation
$\sigma \approx 0.24$. RQ sources show $\overline{\Delta \log \rm
L} \approx -0.06$\ with $\sigma \approx 0.22$, while RL
$\overline{\Delta \log \rm L}\approx -0.07$\ with a slightly
larger $\sigma \approx 0.26$. If we consider separately RQ and RL
sources with the restriction FWHM(\hbbc)$\ge$ 4000 \kms\ (see \S
\ref{rqrlmass}), we find $\overline{\Delta \log L} \approx -0.05$\
(RQ) and $\overline{\Delta \log \rm L} \approx -0.01$\ (RL). It
is worth noting that the scatter is due to a minority of
bad-behaving data points ($\approx$20\%). If they are removed,
the standard deviation becomes $\sigma \approx 0.1$\ in all
cases, with systematic differences  always $\la$0.05.


\section{Mass - Luminosity  Diagram}

Fig.\ref{sample} shows a plot of  \mb\  vs. \Mbh\ for our
combined sample which covers an absolute B magnitude  range $-20
\la$ \mb $\la -27$\ and an estimated black hole mass range $10^7
\la$ M $\la 10^{10} $ (M is everywhere given in \Ms). The data
show a rather well-defined range in L/M with almost all sources
lying between $0.02 \la \rm L/L_{\rm Edd} \la 1.00$.   RQ sources
show evidence for significant Malmquist bias while RL sources
show the opposite trend probably related to a bias towards
selecting higher luminosity  core-dominated sources likely to be
beamed.


 \begin{centering}
 \begin{table*}
 \caption{  RQ/RL sample averages and medians for redshift,
 apparent V magnitude (\mv), absolute B magnitude (\mb), BH mass $(\rm M)$,
and luminosity-to-mass ratio. Average values are given with sample standard
deviations, median values with first and third quartile values. P$_{\rm KS}$ is
the probability that the two distributions are randomly drawn from the same
parent population using Kolmogorov-Smirnov tests. }

\begin{tabular}{lcrrrrrrrrrrrrrrrrrrrrrr}
\hline
& N & \multicolumn{3}{c}{z} && \multicolumn{3}{c}{\mv} &&\multicolumn{3}{c}{\mb}  \\
\cline{3-5} \cline{7-9} \cline{11-13}
\\
 && Aver.$\pm \sigma$  & Med.$_{25\%}^{75\%}$   & $P_{\rm KS}$ && Aver.$\pm \sigma$  & Med.$_{25\%}^{75\%}$   & $P_{\rm KS}$ && Aver.$\pm \sigma$  & Med.$_{25\%}^{75\%}$   & $P_{\rm KS}$\\
\hline
& & & & \\
RQ   & 202  & 0.13$\pm$0.12  & 0.089$_{0.045}^{0.167}$ & $10^{-16}$ && 15.5$\pm$1.1  & 15.5$_{14.7}^{16.2}$    & 1.5 10$^{-4}$ && -23.0$\pm$1.8    & -23.0$_{-24.1}^{-21.6}$ & $10^{-14}$  \\
RL   &  76  & 0.35$\pm$0.21  & 0.334$_{0.200}^{0.530}$ & && 16.1$\pm$0.9  & 16.1$_{15.6}^{16.5}$    & && -24.8$\pm$1.8    & -25.3$_{-26.2}^{-24.2}$ & \\

\hline
%
\end{tabular}
\vskip 0.3cm

\begin{tabular}{lcrrrrrrrrrrrrrrrrrrrrrr}
\hline & \multicolumn{3}{c}{\Mbh}  &&\multicolumn{3}{c}{L/M}\\
\cline{2-4} \cline{6-8}
\\
& Aver.$\pm \sigma$  & Med.$_{25\%}^{75\%}$   & $P_{\rm KS}$  && Aver.$\pm
\sigma$    & Med.$_{25\%}^{75\%}$   & $P_{\rm KS}$\\
    \\   \hline
%
         & & & \\
RQ       & 8.21$\pm$0.76     & 8.21$_{7.65}^{8.79}$   & $10^{-16}$ && 3.89$\pm$0.53   & 3.95$_{3.49}^{4.24}$    & 0.0017  \\
RL       & 9.23$\pm$0.60     & 9.24$_{8.92}^{9.67}$   & && 3.65$\pm$0.47   & 3.63$_{3.37}^{4.03}$    & \\

\hline
\end{tabular}
\label{tabAB}
\end{table*}
\end{centering}


\begin{figure}
 \mbox{}
 \vspace{4.5cm}
 \includegraphics{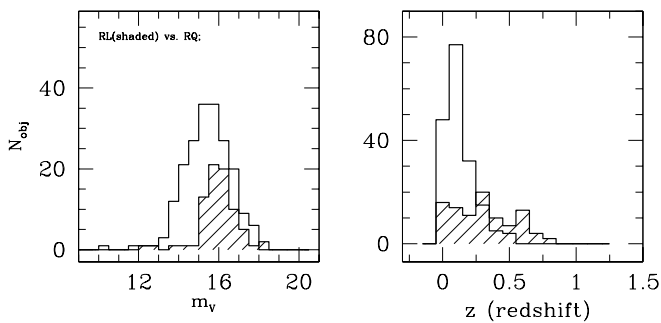}
 \caption{Distribution of apparent V magnitude (left panels) and redshift $z$\  (right
 panels) for our 278 AGN sample. Shaded:      RL;  Unshaded:    RQ. The left panel
 shows that, in our sample, RL sources are on average less bright than RQ sources
 while the right  panel
 suggests a distribution favoring higher $z$ for      RL AGN.  } \label{fig:sample0}
\end{figure}

\begin{figure}
 \mbox{}
 \vspace{8.5cm}
\includegraphics{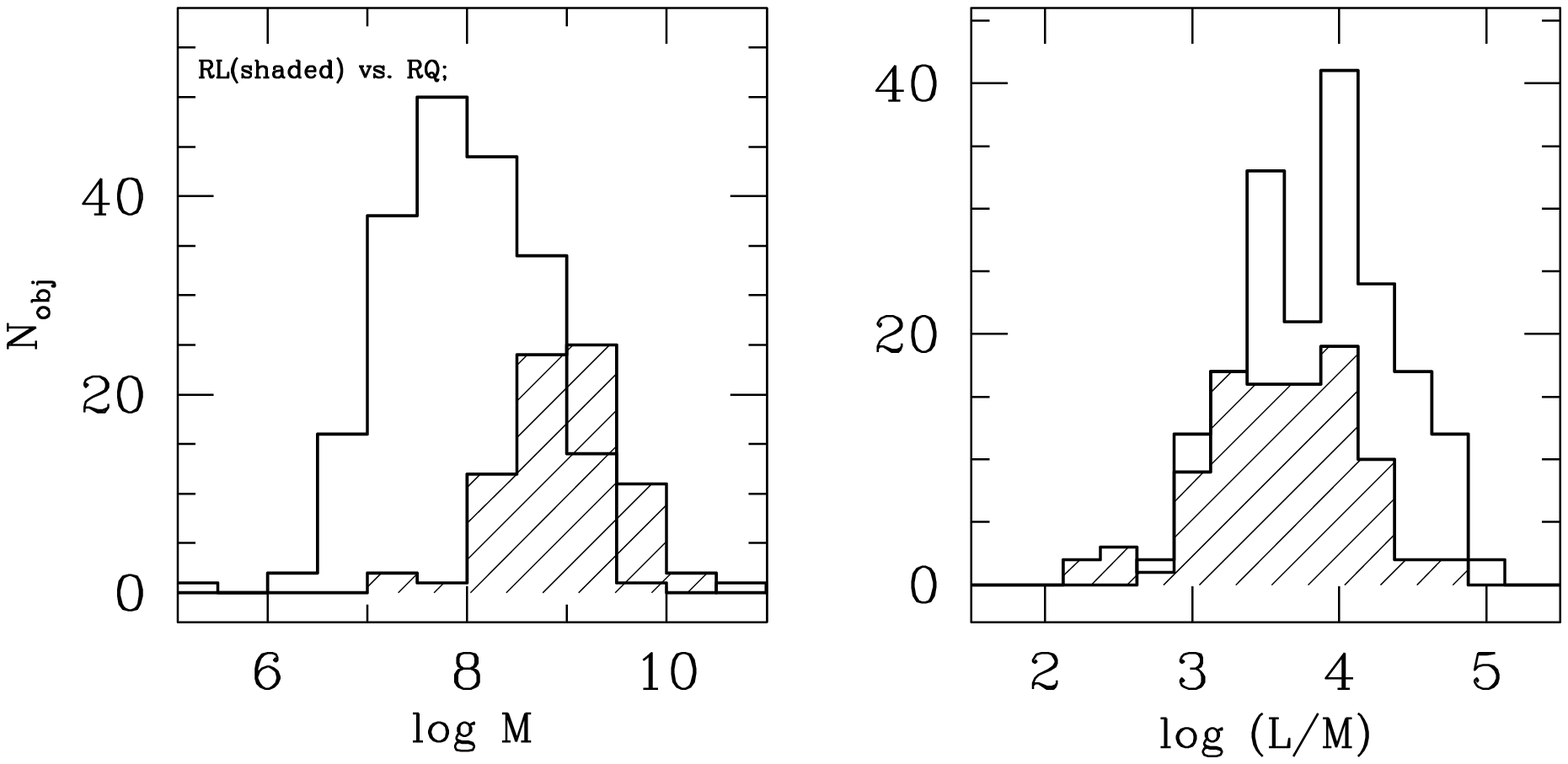}
\includegraphics{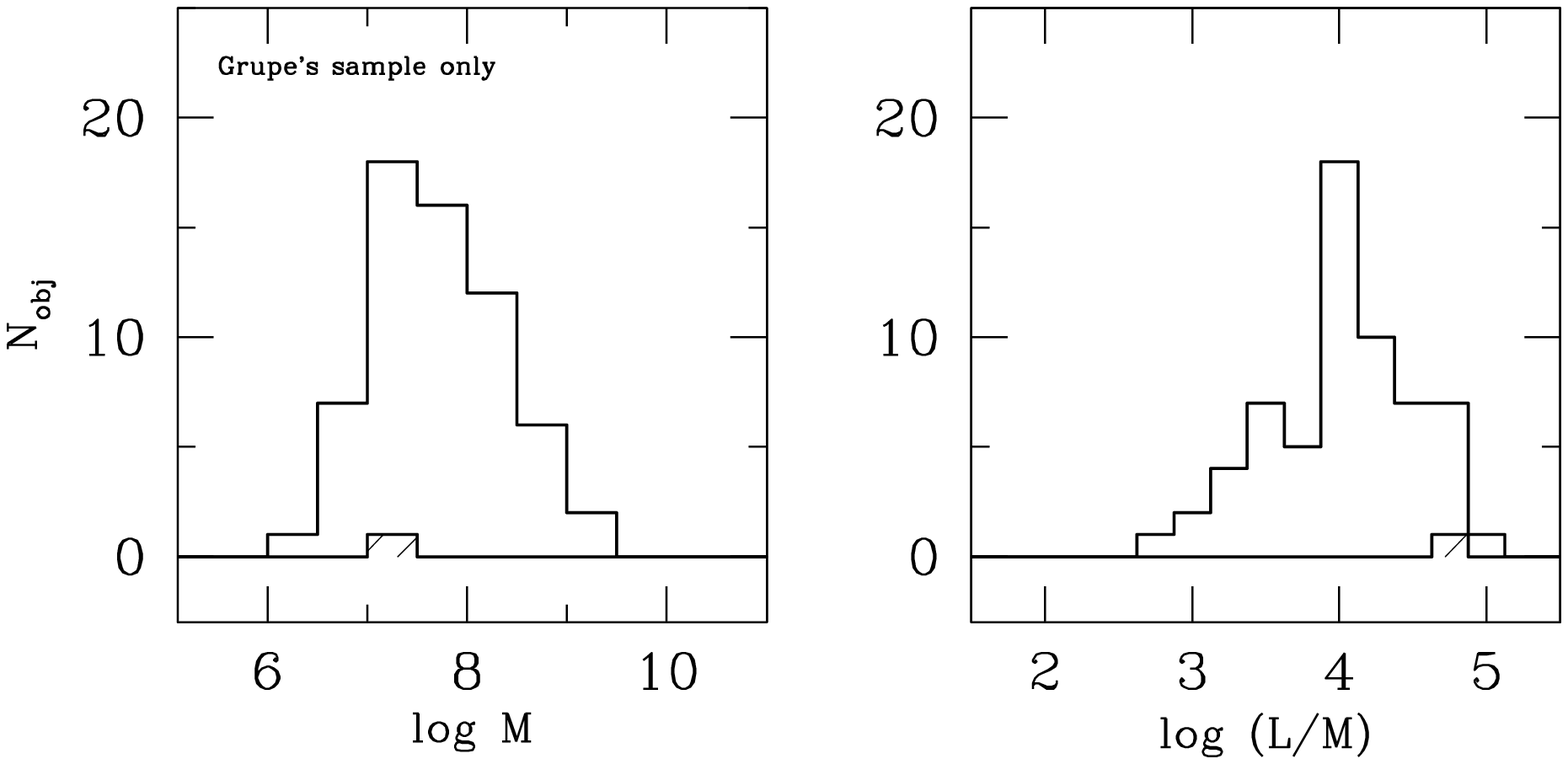} \
 \caption{Distribution of M (left panels) and L/M ratio (right
 panels) for our sample (upper panels) and for the sample by Grupe et al. 1999 only.
 Shaded:      RL;  Unshaded:    RQ. The left panel
 shows that RL sources shows larger M than RQ while the right  panel
 suggests a largely overlapping L/M range for RQ and RL AGN, even if the
 largest L/M sources are almost all RQ.  } \label{fig:sample2}
\end{figure}

\begin{figure}
 \mbox{}
 \vspace{4.5cm}
 \includegraphics{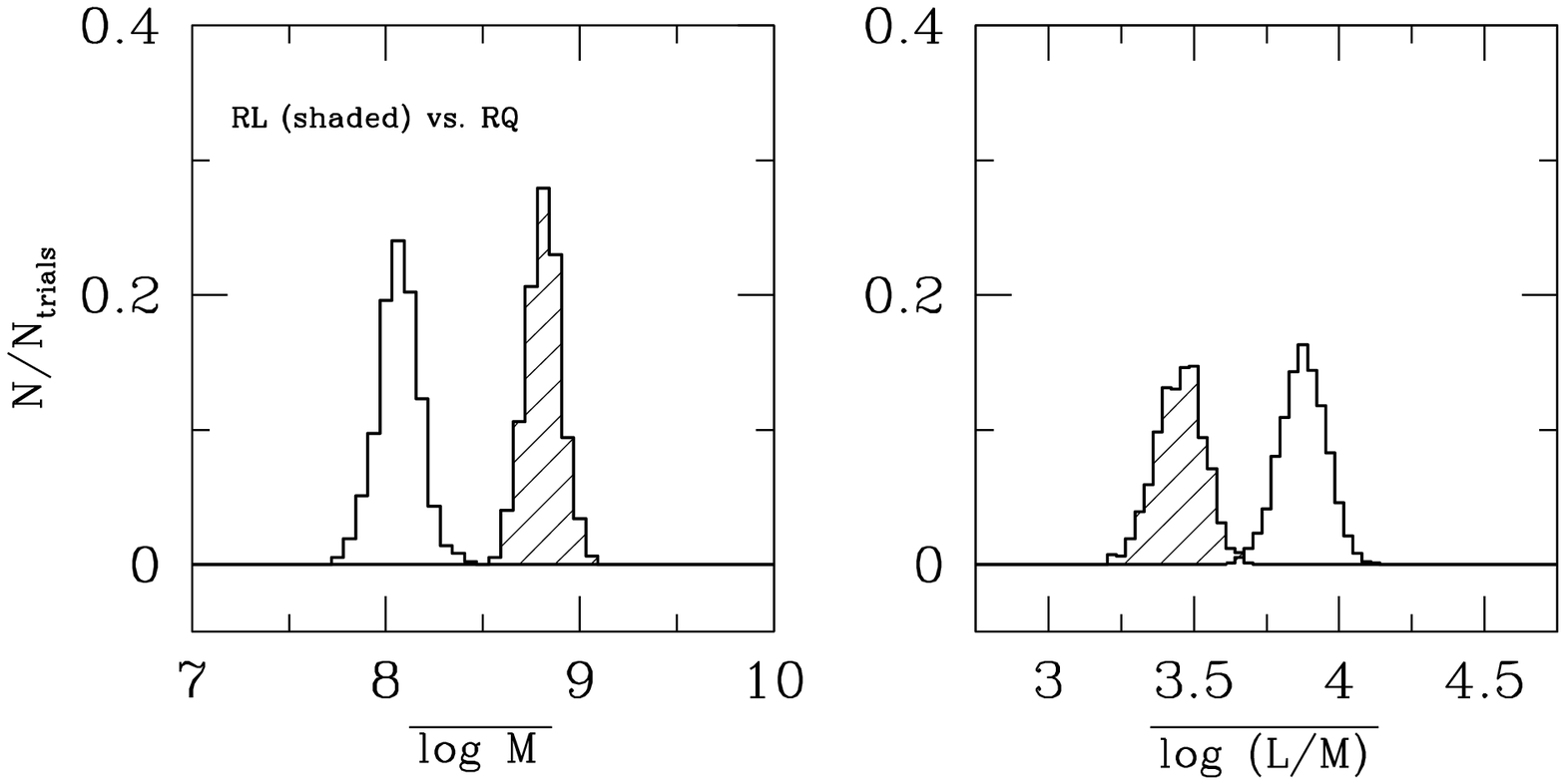}
 \caption{Results of bootstrap simulations.
 Distributions of average M (left) and L/M (right) for
 1000 pseudo samples of 40  RL  (shaded) and RQ objects
 with matching apparent magnitude and
 redshift distribution.    Bootstrap simulations suggest that there
 are systematic differences in the sense that RL sources have large M
 and  lower  L/M ratios than RQ AGN.
 }
 \label{fig:zmatch}
 \end{figure}

\section{Radio Loud and Radio Quiet AGN \label{rqrl}}


Table~\ref{tabAB} presents average and median values of relevant
properties for the  RL and RQ subsamples. Kolmogorov-Smirnov
tests indicate that RL and RQ samples differ significantly in
redshift, \mv\ and absolute B magnitude distribution. The
distribution of the computed parameters \Mbh\ and L/M are also
significantly different for RQ and RL samples.

RL sources are somewhat over-represented in the M03 sample, but not in the
M03+G99 sample. Any over-representation can be properly quantified if we
compare general population expectation for the number ratio between the
number of RL sources and the total number of sources ($\rm f_R
=$N(RL)/[N(RL)+N(RQ)]) in the \mb\ range covered by our sample. Weighting
the luminosity-dependent $\rm f_R$\ value from La Franca et al. (1994) over
the \mb\ distribution of our sample (see Fig. 2 of M03) i.e., within $-27
<$ \mb\ $ < -23$, we obtain   $\rm f_R \approx$ 25\%\ versus $\rm f_R
\approx$ 39\%\ in M03 and $\rm f_R \approx$ 26\%\ in M03+G99.


More cumbersome are the biases shown in Fig. \ref{fig:sample0}:
RL sources tend to be fainter and more distant than RQ ones, as
well as intrinsically more luminous. Biases that are affecting  RL
sources are not clear. RL core-dominated sources are
over-represented  in M03. This will contribute to bias the RL
sample to higher L and $z$\ values (see Sulentic et al. 2003 for
a more thorough discussion).

\subsection{Bootstrap Simulations \label{boo}}

Are differences in redshift and luminosity distributions the
cause of the inferred M and L/M differences between  RQ and RL
sources? We can use most of our sources in creating better
matched subsamples out of our M03+G99 sample (bootstrap
techniques, i.e. Efron \& Tibshirani, 1993). The most important
parameter appears to be redshift (which implies a strong bound on
the apparent magnitude). We constructed pseudo-samples of  RQ and
RL sources {\em with similar redshift and apparent magnitude
distributions}. More precisely: (1) we randomly selected two
pseudo-samples with the same $z$\ and \mv\ distributions,
indistinguishable within  a 2$\sigma$ confidence limit, (2) we
computed average $\log$M  and $\log$L/M values for each
pseudo-sample, (3) we repeated the selection $\sim$1000 times,
(4) we computed the distribution of average $\log$M and $\log$
L/M values for the pseudo-samples (the {\em bootstrapped}
samples), (5) we derived the expectation values for
$<\overline{\log{\rm M}}>$\ and $<\overline{\log{\rm L/M}}>$\ as
the medians of the distributions for the bootstrapped samples,
(6) we estimated the significance level of any difference  from
the average values and the dispersion of the distributions. While
cannot eliminate biases from our sample, we can try to apply the
same bias to both populations.

\subsection{Matching $z$\ and \mv\ Distributions}  The bootstrap procedure
makes both the M  and L/M differences stronger (see the upper panels of
Fig.\ref{fig:sample2} and Fig.\ref{fig:zmatch}): the median values are
$<\overline{\log \rm M}(\rm RQ)> \approx 8.07 \pm 0.07$\ vs.
$<\overline{\log \rm M}(\rm RL)> \approx 8.81 \pm 0.06$, and
$<\overline{\log \rm L/M}(\rm RQ)> \approx 3.88 \pm 0.05$\ vs.
$<\overline{\log \rm L/M}(\rm RL)> \approx 3.45 \pm 0.06$. The results can
be seen in Fig.\ref{fig:sample2} (the original sample) and
Fig.\ref{fig:zmatch} (after bootstrap simulations).  Since matching \mv\
and $z$\ is equivalent to a match in L, it follows that the two plots are
not independent, and that systematically larger FWHM(\hbbc) drives a larger
mass estimate for RL sources.

\subsection{Same $z$\ and \mv\ Distribution with a Narrow Mass Range \label{rqrlmass}}
\begin{figure}
  \mbox{}
  \vspace{4.5cm}
  \includegraphics{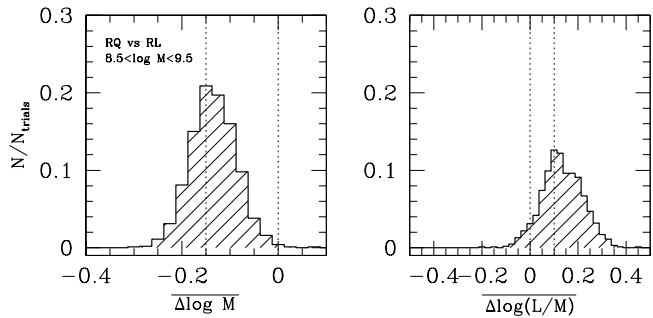}
  \caption[]{Distribution of  $\Delta \overline{\log \rm M}$\ =
  $\overline{\log \rm M}$(RQ) -- $\overline{\log \rm M}$(RL)
  (left panel) and $\Delta \overline{\log \rm L/M}$ = $\overline{\log \rm L/M}$(RQ)
   -- $\overline{\log \rm L/M}$(RL) (right panel)
  for $\sim$1000 pseudo samples of RL and RQ sources.
  \mv\ and  $z$\ distributions were matched, and a restriction to the
  \Mbh\ was applied:  $8.5\le \log$\Mbh$\le9.5$ (48 sources).
   RL and RQ  quasars
  have almost the same L/M ratio distributions.
  }
 \label{fig5}
\end{figure}

\begin{figure}
 \mbox{}
 \vspace{5.5cm}
 \includegraphics{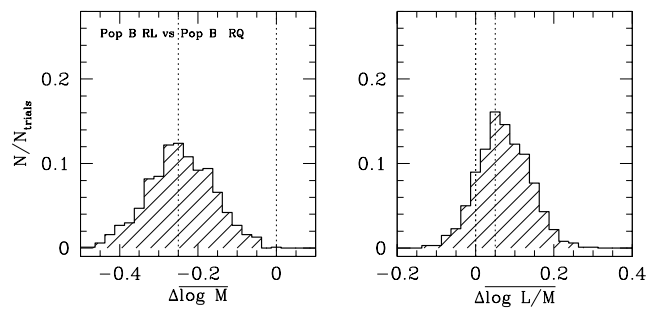}
 \caption[]{Distribution of $\Delta \overline{\log M}$ (left panels)
 and $\Delta \overline{\log L/M}$\ (right panels)
 for $\sim$1000 pseudo samples of 40  objects with FWHM(\hbbc)$\ge$4000\kms\  but separated on the basis of radio loudness.
 It points that  RL and RQ  \PopB\  are not distinguishable on the basis of \Mbh\  or L/M. }
 \label{fig:boo}
\end{figure}

To avoid sample biases, and to  assess the reality of any L/M
systematic difference, we restricted our attention to sources
within a narrow mass range. At the same time, we retained the
above conditions on the $z$\ and \mv\ distribution. The mass
range was chosen so that we had the maximum possible number of
sources: $8.5 \la \log$ M~ $\la 9.5$\ with 48 sources (Fig.
\ref{fig5}). No systematic difference in $\log \rm L/M$ ($\la
0.2$) is found. In a comparison of RQ and RL  sources with the
additional condition FWHM(\hbbc) $\ga$ 4000 \kms, and  matching
the $z$\ and \mv\ distributions, we find again no systematic
difference in either M or L/M distributions (see Fig.
\ref{fig:boo}; note that the difference can be statistically
significant, but it is of the order of unceratainty in M and L/M
determination and therefore too small to be of relevance). The
similarity of the parameter space occupation for RQ and RL
sources with FWHM(\hbbc) $\ga$ 4000 \kms\ carries into a
similarity in derived L/M and M. In other words, there is a large
RQ population whose M and L/M values are similar to the ones of
RL sources.

\subsection{Population A and B: Understanding  Biases \label{popab}}

Sulentic et al. (2000a) introduced the concept of Population A
[FWHM(\hbbc) $\la$ 4000 \kms] and Population B
[FWHM(\hbbc)$\ga$4000 \kms] as fundamentally related to the BLR
structural properties (Sulentic et al. 2000a,b). RQ sources
dominate  Pop. A (88 \%\ of the total)  and are about
$\frac{1}{2}$ of Pop. B sources. Most RL sources ($\approx$75\%)
are Population B. Thus to a large extent a RL vs. RQ comparison
is a Pop. A  RQ vs. Pop. B RL comparison (as it was in the early
study on \civ\ by Marziani et al. 1996).

The validity of the previous results on M and L/M difference
depends on (1) the fraction of \PopA\ RQ sources at low $z$, and
on (2) whether most RL sources at $z<1$ are \PopB. A proper
assessment of the Pop. A/Pop. B ratio for RQ samples  would imply
a thorough analysis of the discovery biases affecting major
surveys (see e.g., Oshlak et al. 2002). This goes far beyond the
aim of the present paper.

Actually, we know that we have a bias favoring \PopA\ RQ sources in the G99
sample. If we apply the same bootstrap analysis to the M03 sample only, we
obtain a lower fraction of Pop. A sources ($\approx$ 45\%), but we reach the
same qualitative conclusions. Therefore a significant bias may arise only if
we {\em miss} a large fraction of \PopB\ RQ sources ($>$20\%). This does not
seem the case. An analysis of the SDSS shows that NLSy1s should be $\approx$
15\%\ of all low-$z$\ AGN (Williams et al. 2002). The ratio between NLSy1s and
the rest of Pop. A (i.e., sources satisfying the criterion 2000 \kms $\la
$FHWM(\hbbc)$\la$ 4000 \kms) is $\approx \frac{1}{3}$\  (estimated on the
whole M03 sample, on a restriction to PG sources). This implies that \PopA\
should be $\approx$ 60\%\ of all low $z$\ AGN. We note that: (a) in M03+G99,
Pop. A sources account for $\approx$52\% of all sources; (b) the ratio of
NLSy1s to all RQ sources is $\approx$15--20\% if we consider the M03 and
M03+G99 independently, in line with the SDSS findings. Therefore, if there is
a sample bias favoring one of the two populations, it is not strong.


Regarding the second issue, it has been possible to show that Fanaroff-Riley
II quasars belong exclusively to Pop. B.  RL sources with narrower lines [RL
with FWHM(\hbbc)$\la$4000 \kms] are predominantly core-dominated (CD) sources,
and therefore likely to be a less frequent, beamed population of preferentially
aligned sources in a FRII/CD unification scenario (Sulentic et al. 2003).

Even if the amplitude of the L/M and M systematic difference will of course
depend on sample definition, such difference is unlikely to be caused by
selection effects, and should be considered real. This is in line with
recent research pointing toward systematic differences in the mass function
of RQ and RL AGN (as discussed in  \S \ref{disc:rqrl}).

%
%
%

\section{\hbbc\ Profile Shape}

\subsection{The Influence of \Mbh\ and L/M}

\begin{figure}
 \mbox{}
 \vspace{8.5cm}
 \includegraphics{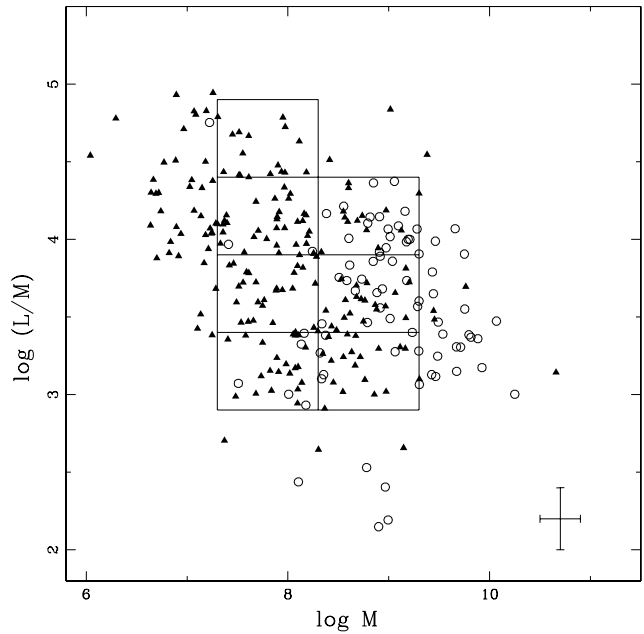}
 \caption[]{ Binning of our sample in the  L/M vs. \Mbh\
parameter space for the median spectra computations. Symbols are as in
Figure 1.
   }
 \label{bins}
\end{figure}

\begin{figure}
 \mbox{}
 \vspace{14.0cm}
  \includegraphics{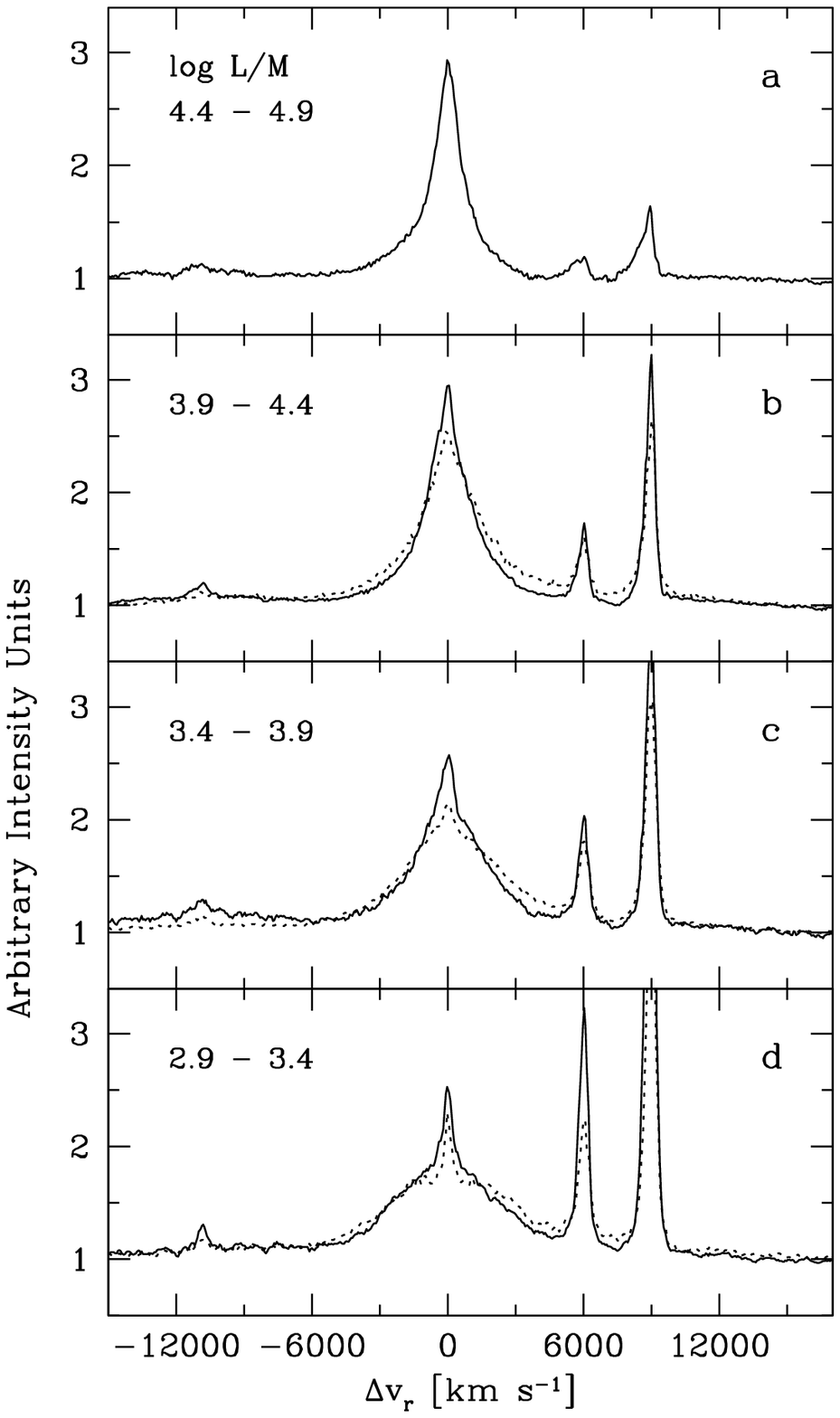}
 \caption[]{ \hb\ line profiles after \feii\ subtraction.
    From top to bottom  the plots correspond to:
    {\bf (a) } $\log$L/M  4.4 - 4.9;
    {\bf (b) } $\log$L/M  3.9 - 4.4;
    {\bf (c) } $\log$L/M  3.4 - 3.9;
    {\bf (d) } $\log$L/M  2.9 - 3.4. Note that $\log$ L/M
    $\approx$4.53 corresponds to $\rm L/L_{\rm Edd} \approx 1.$
    Solid lines correspond to $7.3\le \log$\Mbh$\le 8.3$,
    and dashed lines to $8.3 < \log$\Mbh $\le 9.3$.
   }
 \label{profiles}
\end{figure}


We generated median  \hbbc\ profiles in bins covering narrow ranges of
derived \Mbh\ and \lm. The adopted binning is shown in Fig. \ref{bins}. The
median spectra (normalized to the local continuum) are plotted in Fig.
\ref{profiles}. Table  2 provides, for every bin, average measures of the
\hbbc\ profile centroid at 1/4 and 1/2 peak intensity, the best fit to the
profile shape, and the intensity ratio (C2/C1) in case when two components
have been used.

\begin{centering}
\begin{table*}
\caption{\hbbc\ line centroids at 1/4 and 1/2 fractional intensity as a
function of $\log {\rm L/M}$. For each interval of $\log {\rm L/M}$\ the
table reports the \hbbc\ best fitting function (FF; either Lorentzian (L)
or Gaussian (G)), and their peak position and FWHM. In case two component
are needed for the best fit, their intensity ratio (C2/C1)
is reported in the last column.  }
\begin{tabular}{r  rrrrr  rrrrrrrr}
\hline\hline $\log {\rm L/M}$    &    & c(1/4) & c(1/2) & FWHM &
\multicolumn{7}{c}{Best Fit}  \\  \cline{6-13} &&&&&
\multicolumn{3}{c}{1$^{st}$ Component} && \multicolumn{3}{c}{2$^{nd}$\
Component} & C2/C1 \\ \cline{6-8} \cline{10-12}
&&&&& FF & Peak & FWHM && FF & Peak & FWHM &&\\
    & & [\kms]   & [\kms]  & [\kms]  & & [\kms] & [\kms] && &[\kms] & [\kms]        \\
        \hline
    & & & & & & & & & \\
    & & \multicolumn{2}{c}{ $7.5\le \log$\Mbh$\le 8.5$ } & & & & \\
    & & & & & & & & & & & \\
4.4$-$4.9 &      & $-170^{+290}_{-390}$ & $-55^{+120}_{-110}$  & 1410$\pm$110 & L & 50 & 1400 && G & -1850 & 2100 & 0.09 \\
3.9$-$4.4 &      & $+140^{+400}_{-390}$& $+80^{+160}_{-160}$ & 2310$\pm160$ & L   & -5 & 2350 && ... & ... & ... & ...  \\
3.5$-$3.9 &      & $+140^{+500}_{-540}$ & $+200^{+240}_{-240}$ & 3200$\pm240$ & L & 50 & 3500 && ... & ... & ... & ...   \\
2.9$-$3.4 &      & $+400^{+780}_{-600}$ & $-85^{+310}_{-310}$ & 5490$\pm310$ & G & -320 & 4000 && G & +1400 & 10000 & 0.73 \\
          & & & & & & & & & & \\
&& \multicolumn{2}{c}{ $8.5 <\log$\Mbh$\le 9.5$ } & \\
          & & & & & & & & & & & \\
3.9$-$4.4 &  & $+330^{+620}_{-560}$ & $+250^{+250}_{-240}$ & 3100$\pm250$    & L & +1 & 3000 && G  & +4000 & 7600 & 0.12  \\
3.5$-$3.9 &  & $+915^{+1100}_{-790}$ & $+440^{+320}_{-320}$ & 4900$\pm320$ & G & -20 & 2900 && G & +1050 & 8700  & 1.7  \\
2.9$-$3.4 &  & $+2320^{+1060}_{-1090}$ & $+314^{+400}_{-380}$ & 7000$\pm$400 & G & -70 & 4900 && G & +2100 & 12600 & 1.09    \\
\hline\hline
\end{tabular}
\label{tabCen}
\end{table*}
\end{centering}

\begin{figure}
 \mbox{}
 \vspace{13.0cm}
  \includegraphics{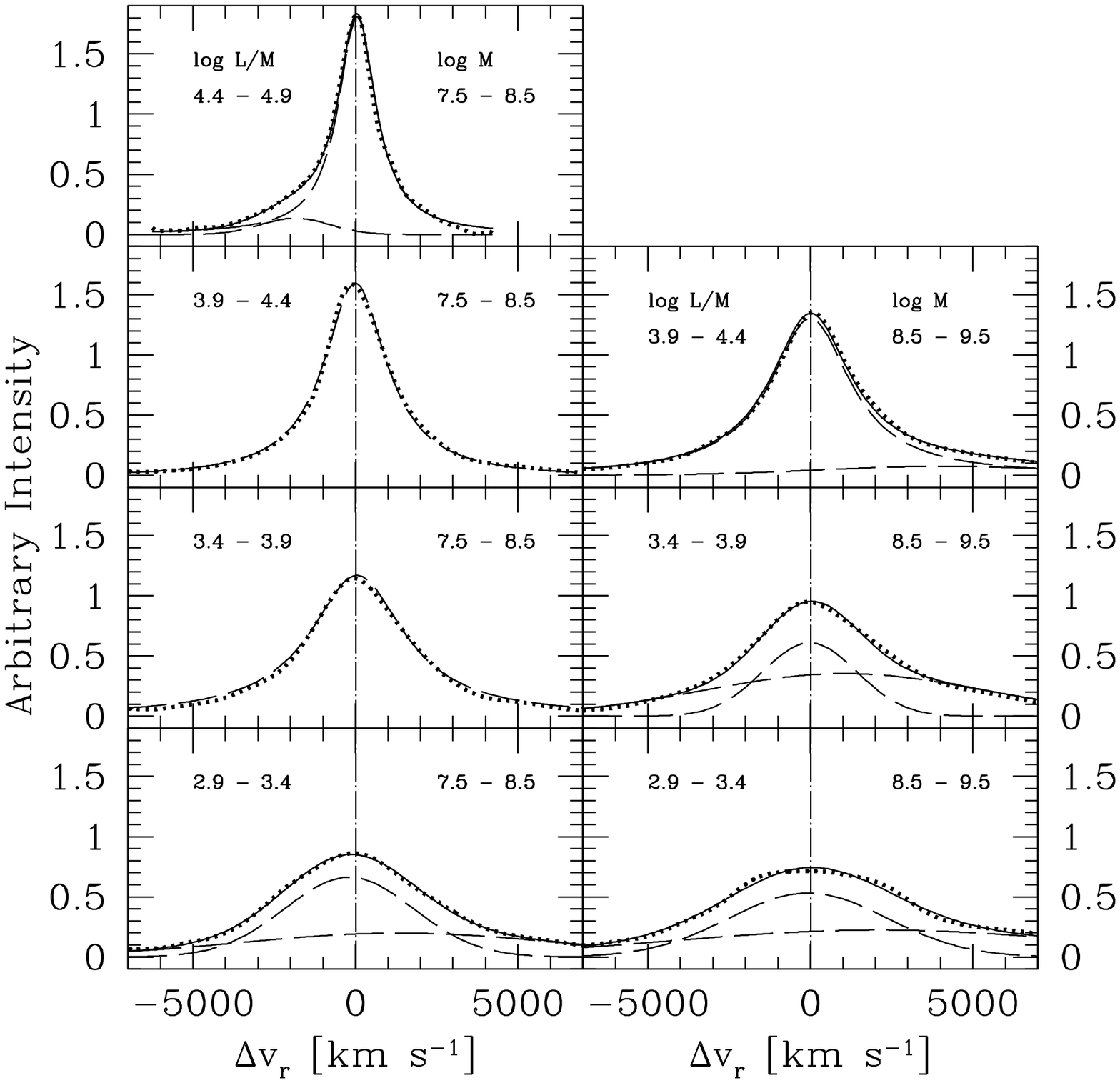}
 \caption[]{\hbbc\ line profile fitting after continuum, \feii\ and \hbnc\ subtraction
 for the intervals in log L/M and M   defined in Table 2.
 The ``cleaned" \hbbc\ (dotted line) is shown along with single fitting components (dashed
 lines) and with the resulting best fit (thin solid line).
   }
 \label{bestfits}
\end{figure}

\begin{figure}
 \mbox{}
 \vspace{6.0cm}
  \includegraphics{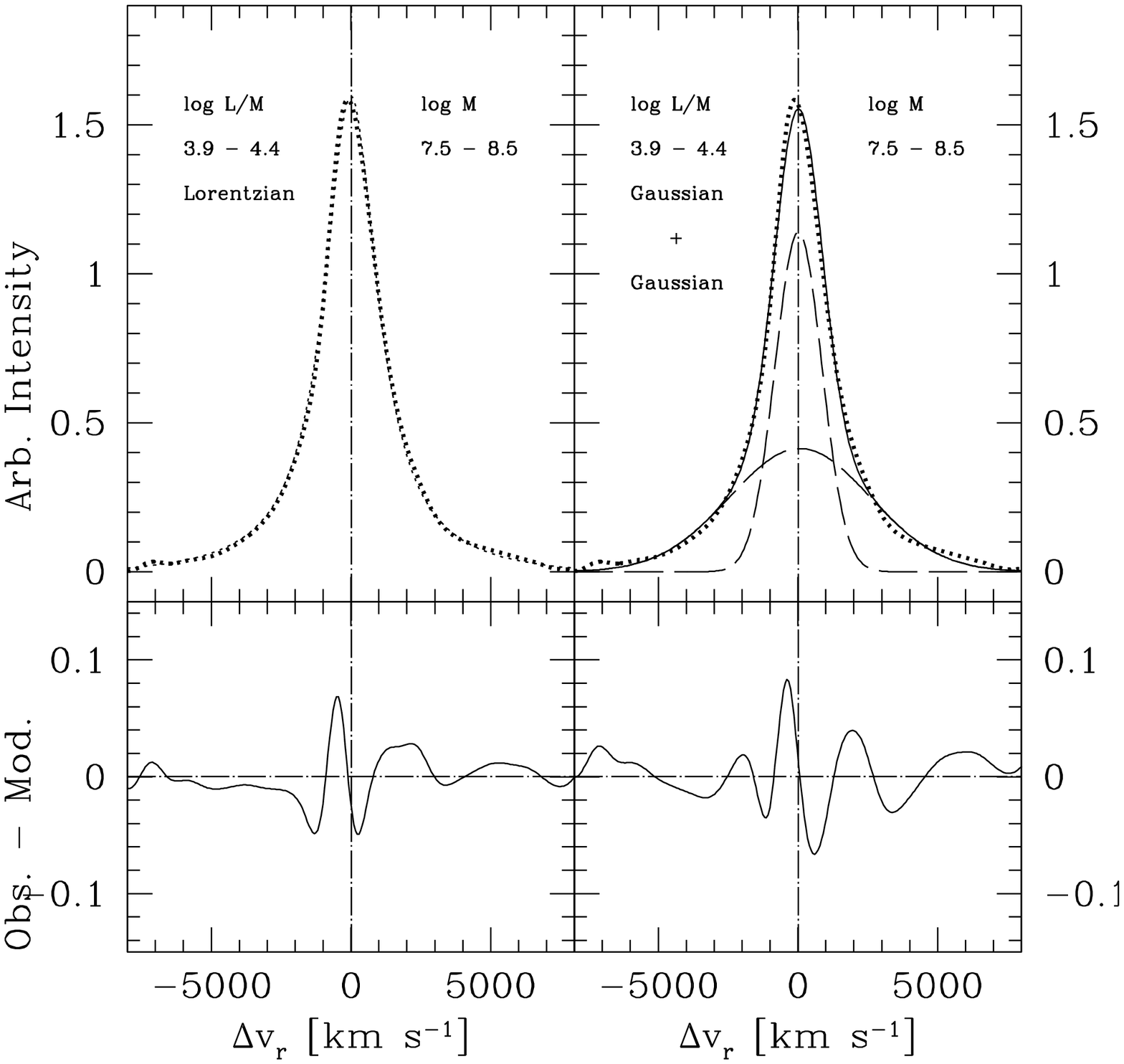}
 \caption[]{Example of \hbbc\ line profile fitting with different functional forms.
The upper panes show the \hbbc\ (dotted lines)  fit by a Lorentzian (left,
best fit  identical to Fig. \ref{bestfits}) and to a double Gaussian. The lower
panel show residuals for both cases. The Lorentzian fit is significantly
better, yielding to a decrease in $\chi^2$\ by a factor $\approx 1.5$. This is
the most doubtful case among the ones considered in this study; in all other
cases a Lorentzian/Gaussian exchange from best fits yields  a $\chi^2$\
worsening by a factor $\approx 2$.   }
 \label{residuals}
\end{figure}

The median \hbbc\ profile computed for each bin (see Fig. \ref{bestfits} and
\ref{residuals})  suggests that:

\begin{itemize}

\item  a Lorentz function provides good fits to median  profiles in bins for which
$3.9 < \log \rm L/M < 4.4$, and for   $3.4 < \log\rm L/M < 3.9$
and $7.5 < \log\rm M < 8.5$. Profile models employing Gaussian
functions yield poorer fits with significantly larger $\chi^2$.
The slight redward asymmetry in the bin  $3.9 < \log\rm L/M <
4.4$ and $8.5 < \log\rm L/M < 9.5$\ can be modeled by adding a
weak, redshifted Gaussian component C2 with C2/C1 $\approx$ 0.1.
Similar considerations apply to the bin $4.4 < \log\rm L/M < 4.9$
and $7.5 < \log\rm M < 8.5$\ (i.e., largest L/M, possibly
super-Eddington, and lowest M) profile: a slight blueward
asymmetry visible in the median profile can be modeled as  an
additional, blueshifted component contributing $\sim 0.1$ of the
total line emission.

\item  If $\log \rm L/M < 3.4$, profiles are better fit with double Gaussian models
than  using a Lorentz core + broader Gaussian component The latter
model is favored if the narrow component of \hb\ is {\em not}
subtracted (e.g. Oshlack et al. 2002, but lack of narrow
component subtraction will also lead to an underestimate of M.

\item  The ``red shelf" or redshifted Gaussian component (we
refer to it as very broad line region (VBLR) component) appears to be
strongly influenced by M, i.e. larger M sources show a more prominent VBLR
component, as well as by L/M (this is especially evident from the values for
8.5$< \log \rm M \le 9.5$\ reported in Table 2). The M dependence is easily
seen by comparing the solid (lower mass) and overlayed dashed (higher mass)
profiles in each frame of Figure \ref{profiles}. Table 3 shows that c(1/4)
measures are significantly redshifted for the bins corresponding to larger
M.

\end{itemize}

Best fits to \hbbc\  and individual line components are shown in Fig.
\ref{bestfits}. Fig. \ref{residuals} illustrates a case in which the exchange
from a Lorentzian to a Gaussian significantly worsen  residuals. The case
shown is the one where fits provides more similar results. In general
exchanging functional forms from the best fitting ones imply  larger
differences and a $\chi^2$\ worsening by a factor of $\approx$ 2.

\subsection{Influence of Radio Loudness on \hbbc}

\begin{figure*}
 \mbox{}
 \vspace{7.0cm}
 \includegraphics{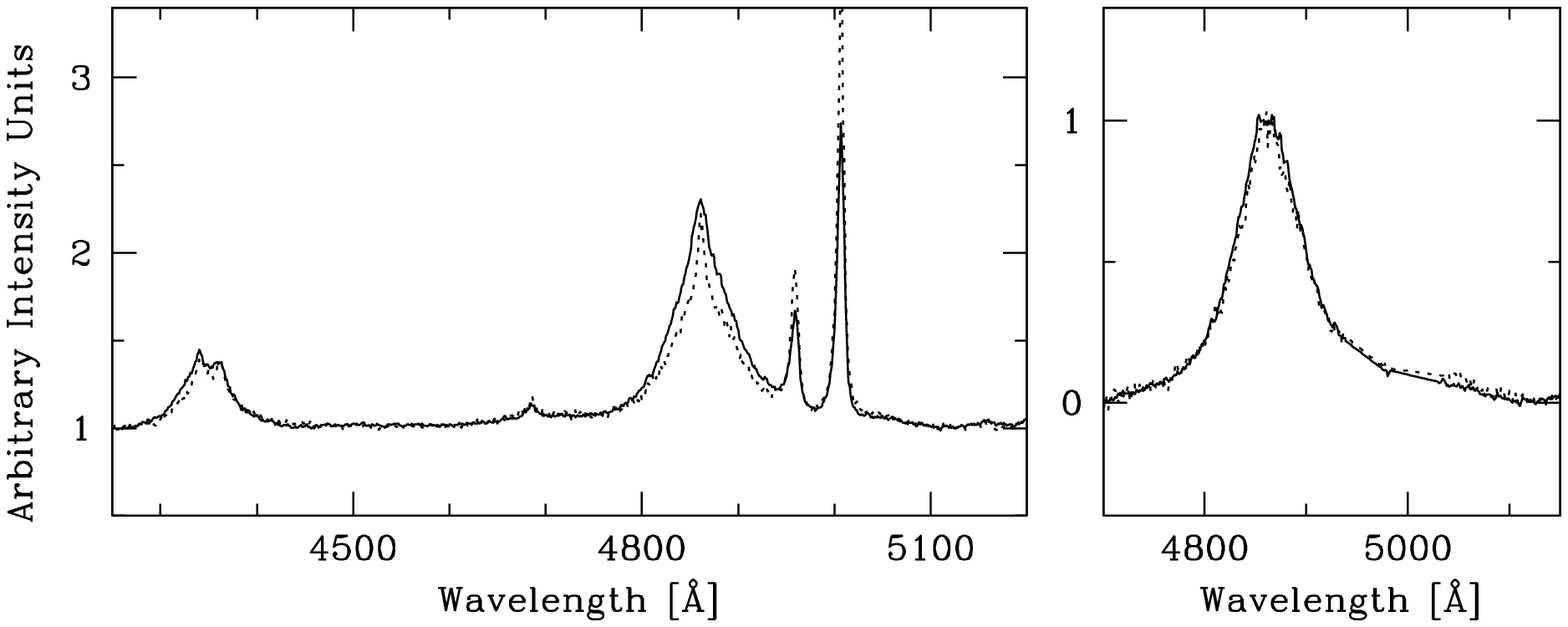}
 \caption{Comparison of the RQ and RL objects in the mass interval
   $8.5<\log$\Mbh$<9.5$ and any L/M ratio
   (N$_{RQ}=$56, N$_{RL}=$36).
   {\bf (leff)  } Comparison of the composite spectra of RQ and RL
      (after subtraction of the \feiiopt\ template).
      W(\hbbc) is  20\% higher in RQ,  W(\oiii) is  35\% higher in RL.
  {\bf (right) } cleaned and scaled \hbbc\ profile.
   The \hbbc\  profiles  of RQ and RL are almost identical.
   }
 \label{profR}
\end{figure*}

Is the \hbbc\ profile shape influenced by radio loudness? To answer  this
question we generated the \hbbc\ profile  (see Fig. \ref{profR}) for the
interval $8.5 < \log$ M $ < 9.5$ and {\em any} L/M  for RQ and RL separately
(N$_{\rm RQ}= $ 56, N$_{\rm RL}= $ 36). Fig. \ref{profR} shows that the
composite spectra of RL and RQ objects are different: the narrow lines
(\oiiiopt\ and \hbnc) are stronger and W(\hbbc) is lower in RL objects. At the
same time, the \hbbc\ profiles remain indistinguishable  within our S/N limits
(see Fig. \ref{profR}). To further check this result we additionally generated
the composite spectra separating  RL and RQ objects in three other M and L/M
ranges: (a) $8.5< \log$M$<9.5$ and 3.9$< \log$L/M $<$ 4.4 (N$_{\rm RQ}=$14,
N$_{\rm RL}=$19);
 (b) $8.5<\log$M$<9.5$ and  3.4$< \log$L/M $<$ 3.9 (N$_{\rm RQ}=$26, N$_{\rm RL}=$13); (c) and
$8.0< \log$ M $<9.0$ and   3.9 $<\log$L/M$<$4.4 (N$_{\rm RQ}=$14, N$_{\rm
RL}=$19). Although the S/N ratio of the composites was lower since fewer
objects were used,  results were  similar: indistinguishable \hbbc\
profiles and centroids, with higher W(\oiiiopt) and lower W(\hbbc) for RL
objects. We conclude that the \hbbc\ line profile shape does not depend
strongly on radio loudness.

\section{The ``Blue Outliers"  \label{bo}}

\subsection{Spectral Properties of \oiiiopt\ ``Blue Outliers"}

In a previous investigation (Zamanov et al. 2002), we identified
seven objects from the M03 sample that showed an \oiiiopt\
blueshift relative to the peak of \hb. The  shift amplitudes were
$\geq$ 250 \kms\ leaving little doubt that we are observing a
significant velocity displacement. We have identified five
additional AGN that show large amplitude blueshifts among the
objects of G99. Some properties of all of these sources are given
in Table \ref{tabBO}. All of the sources show strong \feiiopt\
emission. In the E1 context they show:  (1) \rfe$\ga$0.5, (2)
FWHM(\hb) $<$ 4000 \kms (i.e., they are located in spectral bins
A2 and A3 of Sulentic et al. 2002); (3) a large \civ\ broad
component blueshift (Zamanov et al. 2002). Blue outliers show also
weak \oiiiopt\ emission with W(\oiii)$\la$ 20 \AA, and a median
value of $\approx$6 \AA.

Nine out of our 12 outliers  are formally NLSy1s. In general, the
\oiiiopt\ lines of NLSy1s have a relatively narrow profile with
often, in addition, a second broader, blueshifted component
(V\'eron-Cetty, V\'eron \& Gon\c{c}alves 2001). The blue-shift,
 is   400 -- 1100 \kms\ when the two components can be deblended
 (Zheng et al. 2002). These values correspond to the blue outliers
 range suggesting that the latter are an extreme subset with low W(\oiii) when only
the shifted component is visible.

\subsection{FWHM of \feii\ and \oiiiopt\ lines}

Computation of  M requires that \hbbc\ be separated from any
narrow component so that FWHM(\hbbc) can be properly measured.
The problem is that extreme    sources like the blue outliers show
no profile inflection (reflected in the fact that they are well
fit by a Lorentzian function). The rationale for our methodology
has been detailed in M03, and will not be rediscussed here.  We
just remind that we  normally assume (and find) that the narrow
component of \hb\ shows almost the same shift and width
properties as the \oiiiopt\ lines. However, in blue outliers
W(\oiii) is smaller than in any broad emission line AGN class and
is also blueshifted relative to the peak of \hb. We conclude that
we are not measuring a narrow line component at the Lorentzian
tip of the \hb\ profile. Any narrow component of \hb\ is assumed
to be  blueshifted and lost in the blue wing of the broad line
(even assuming it has the same strength as \oiii\ in blue
outliers it would represent less than 10\%\ of the total line
flux). Subtraction of a narrow \hb\ component would have a
non-negligible effect on the FWHM(\hbbc) measurement which is
needed for the \Mbh\ estimation.

The \feiiq\ feature is a  complex of lines thought to have FWHM similar to
\hbbc. We estimated FWHM(\feiiq) {\em independently} from FWHM(\hbbc) by
\feii\ template fitting using the I~Zw~1 spectrum (of course, only for highest
S/N spectra; see M03 for details). In our sample FWHM(\feiiq) is similar to
FWHM(\hbbc) with few exceptions only (see Bongardo et al. 2002). Estimated
FWHM(\feiiq) values are  given in Table~\ref{tabBO}. No narrow \feiiq\
component has been detected and nor is one expected given the low inferred
density for the narrow line region. In principle, we can use FWHM(\feiiq)
instead of FWHM(\hbbc) in the mass derivations in order to avoid any error
associated with a narrow line component.  Use of FWHM(\feiiq) instead of
FWHM(\hbbc) will yield \Mbh\ values within $\sim$10\% of ones derived from
FWHM(\hbbc).

The correlation of nuclear BH mass with stellar bulge velocity dispersion
$\sigma _\ast$ is now well established in nearby galaxies. Recent results
(Nelson 2000; Boroson 2002) show that a BH mass -- FWHM(\oiii) correlation is
also present but with larger scatter. The FWHM(\oiii) measures given in
Table~\ref{tabBO} provide another way to estimate BH masses. The derived
values are however considerably higher than values calculated  using
FWHM(\hbbc) and source luminosity.  This is not surprising because NLSy1-type
AGN apparently do not follow Nelson's relation (Mathur, Kuraszkiewicz \&
Czerny 2001). If NLSy1s do not follow the relation then blue outliers will
almost certainly show the same lack of agreement.  This does not contradict
our assumption that blueshifted \oiiiopt\ arises in outflowing gas (Zamanov et
al. 2002), possibly associated with a disk wind. The \oiiiopt\ region in blue
outliers may be very compact and its velocity field is not likely to be
dynamically related to the host galaxy stellar bulge. This points to a
limiting W(\oiii) ($\approx$ 20\AA) below which FWHM(\oiiiopt) emission ceases
to be a useful mass estimator.

\subsection{ \mb\  and L/M  Ratio of [OIII] Outliers \label{bo}}

\begin{figure}
\mbox{} \vspace{9.5cm}
  \includegraphics{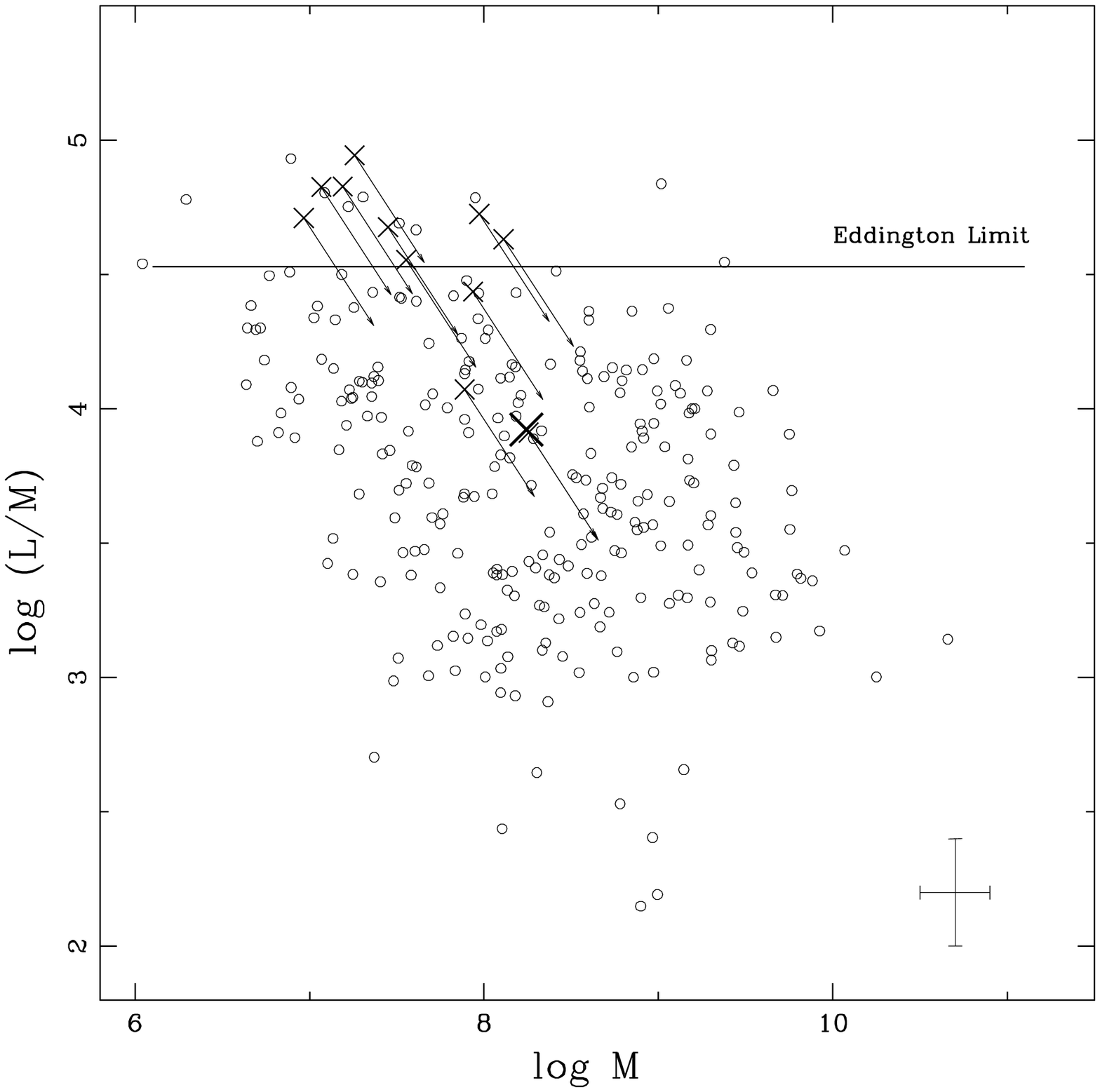}
 \caption[]{Position of blue outliers (crosses) relative to other sources
 in our sample (open circles). The larger cross indicates the position
of PKS 0736+01 - the only known RL ``blue outlier".
 The dashed line indicates the Eddington limit.  Arrows indicate
 displacements of blue outliers if we apply an orientation correction
 $\sim$0.4 to the derived masses.  Blue outliers are among the
 highest L/M sources and remain in the upper part of the diagram even
 after an orientation correction. One should consider that many other sources would move
 in the same direction of the blue outliers if a proper orientation were applied to all sources.
 }
 \label{fig10}
 \end{figure}

All blue outliers lie in the region of E1 thought to be populated by highly
accreting (high \lm) sources (Marziani et al. 2001). Most of them are located
close to the Eddington limit ($\log \rm L/M \approx 4.53$, see
Fig.\ref{fig10}). This is also true if: (1) FWHM(\feiiq) is used instead of
FWHM(\hbbc) to derive M, and (2) an orientation correction is applied.  The
blue outliers may be oriented close to pole-on. Since M $\propto \rm FWHM ^2$,
an inclination correction will move the ``blue outliers" toward higher mass
and lower L/M (see the arrows on Fig.\ref{fig10}). Even with an inclination
correction by a factor $\approx$4, the \lm\ values for blue outliers remain
among the highest observed. We note that: (1) other sources should also be
corrected for orientation especially if low ionization lines are emitted in a
flattened configuration in all  Population A sources, as suggested (Marziani
et al. 1996); (2) some care is needed since we neglected a photometric
correction which is basically unknown. We do not know whether any sort of
continuum beaming  operates for RQ AGN as $i \rightarrow0^\circ$\ (very
similar W(\hbbc) for CD and LD RL sources suggests no strong effect; Sulentic
et al. 2003). Observational errors and uncertainty in the orientation
correction do not allow us to determine whether some blue outliers are really
super-Eddington sources as they appear in Fig.\ref{fig10}.

All blue outliers show estimated masses less than $10^{8.5}\rm M_\odot$. We
tried to determine if the blue outliers large L/M  was a consequence of a
low-mass bias by calculating the \mb\ difference between blue outliers and
other sources in two mass ranges: (1) $7.15 \le \log$ M $\le 8.45$ (where the
``blue outliers" are before the orientation correction) and (2) $ 7.55 \le
\log $ M $\le 8.85$ (where they move after the orientation correction).
Table~\ref{MBofBO} shows that the blue outliers are almost certainly $\sim$ 1
mag brighter than  other sources with similar mass implying 2--3 times higher
L/M ratios.

\begin{centering}
\begin{table*}
\caption{ The spectral properties of the ``blue outliers": radial velocity
difference between top of \hb\ and \oiii, equivalent width and FWHM of
\oiii, FWHM of \feiiq, the equivalent widths ratio of \feiiq\ and \hbbc\
(\rfe).}
\begin{tabular}{lrccccccc}
\hline\hline
  Name     & $\Delta$v  & W[OIII]$\lambda$5007      & FWHM[OIII]     & FWHM(\hb)& FWHM(FeII) & RFe  & \\
       & [\kms]& [ \AA\ ]             & [\kms]  & [\kms]   & [\kms]     & & \\
\hline
  I Zw 1              & -640 & 15.3 & 1440$\pm$120 & 1092  & 1095   & 1.30$\pm$0.1 &\\
  PKS 0736+01$^*$     & -430 & 2.6  &  720$\pm$60  & 3258  & 3560   & 0.70$\pm$0.1 &\\
  PG 0804+761         & -305 & 10.1 &  780$\pm$60  & 3302  & 3067   & 0.42$\pm$0.1 &\\
  PG 1001+291         & -680 & 3.4  &  960$\pm$60  & 1759  & 1588   & 0.71$\pm$0.1 &\\
  PG 1402+261         & -300 & 2.6  &  900$\pm$180 & 1938  & 1957   & 0.73$\pm$0.1 &\\
  PG 1415+451         & -600 & 2.9  &  660$\pm$60  & 2555  & 2870   & 0.66$\pm$0.1 &\\
  PG 1543+489         & -950 & 6.5  &  ---         & 1555  & 1588   & 0.64$\pm$0.1 &\\
  RX J0136.9-3510     & -380 & 6.0  &  900$\pm$60  & 1050  & 1100   & 1.0$\pm$0.1  &\\
  RX J0439.7-4540     & -580 & 5.0  & 1020$\pm$120 & 1020  & 1100   & 0.9$\pm$0.1  &\\
  RX J2217.9-5941     & -330 & 8.1  & 1140$\pm$120 & 1370  & 1588   & 2.0$\pm$0.2  &\\
  RX J2340.6-5329     & -490 & 21.8 &  780$\pm$60  & 1228  & 1588   & 1.3$\pm$0.1  &\\
  MS 2340.9-1511      & -420 & 5.8  &  780$\pm$60  &  969  & 1218   & 1.1$\pm$0.1  &\\

\hline\hline $*$ RL CD source

\label{tabBO}
\end{tabular}

\end{table*}
\end{centering}


\begin{centering}
\begin{table}
\caption{ \mb\ of the ``blue outliers" (as they are given in  V\'eron-Cetty
\& V\'eron catalog without correction for the cosmology), compared with
other objects in the sample with similar mass. P$_{\rm KS}$\ is the
probability that the ``blue outliers" and the other objects in the
indicated mass ranges are randomly drawn from the same parent population
according to Kolmogorov-Smirnov test. }
\begin{tabular}{lrrr}
\hline\hline
                                & Aver.$\pm \sigma$ & Med.$_{25\%}^{75\%}$        & $P_{\rm KS}$  \\
\hline
                                & & & \\
blue outliers                   & $-23.9 \pm 0.8$   & $-23.85^{-23.45}_{-24.30}$  & \\
$ 7.0 \le \log$\Mbh$\le 8.2 $ & $-22.4 \pm 0.8$   & $-22.20^{-21.50}_{-23.40}$  & 0.00035    \\
$ 7.6 \le \log$\Mbh$\le 8.8 $ & $-22.8 \pm 0.8$   & $-22.90^{-21.80}_{-23.90}$  & 0.00600    \\

\hline\hline
\end{tabular}
\label{MBofBO}
\end{table}
\end{centering}

\section{Discussion \label{discuss}}

\subsection{Mass-Luminosity Diagram}

The ranges of \Mbh\ and \lm\ considered in this study are similar
to those in other recent work (Woo \& Urry 2002a; Collin et al.
2002). We find that the Eddington limit defines an approximate
upper boundary to the luminosity distribution (as in Woo \& Urry
2002a), indicating that there are no known low-$z$\ AGN accreting
significantly above the Eddington limit (this depends on the
adopted $\rm H_0$\ value).  Figure \ref{sample} also suggests
that there may be fewer high (than low) mass  sources with \lm\
close to the Eddington limit.  This might be a selection effect
or an indication that galaxies with a high mass BH may be unable
to supply fuel at high L/M. In other words, the upper envelope in
Figure \ref{sample} might be due to the Eddington limit as well
as the ability of the surrounding matter to feed the accretion
flow (see, e.g., Nicastro et al. 2003). The lower envelope of the
luminosity distribution may be due to a selection effect (e.g.
Woo \& Urry 2002a), or it   may indicate that only sources
radiating at $0.01 \la (L/L_{Edd}) \la 1.00 $ exhibit a stable
broad lines region. The latter possibility might be connected
with the presence/absence of an accretion disk wind (see also
Kollatschny \& Bischoff 2002).

\subsection{On the Difference Between Radio Loud and Radio Quiet AGN \label{disc:rqrl}}

Somewhat confusing claims have been recently made on the
difference between RL and RQ sources in terms of M and L/M. Laor
(2000) found that the radio loudness is strongly related to \Mbh.
Woo \& Urry (2002b) concluded that radio loudness does not depend
strongly on M. Lacy et al. (2001) found no evidences for critical
\Mbh\ or L/M ratio that turns on powerful radio jets. Ho (2002)
showed that the dependence of radio loudness on \Mbh\ disappears
altogether when one considers AGN with a broad range of intrinsic
luminosity. At the same time he found that radio loudness seems to
be related to the mass accretion rate. RL and RQ sources are well
separated {\em in terms of} \Mbh\ by the new Eigenvector analysis
of Boroson (2002).

The difference between RQ and RL on Eigenvectors diagram (Boroson 2002) is
probably connected with the inclusion of \oiii\ line parameters. In RL quasars
part of the \oiiiopt\ is coming from extended regions connected with the radio
jet (Wilman, Johnstone \& Crawford 2000). Previous work has shown that RL
activity in low M sources may be physically possible. Our sample contains a RL
NLSy1 that is one of the highest L/M sources, and other sources that radiate
al $\log \rm L/M \approx 4.5$. Our analysis stressed that there are no
appreciable effects (within the limits set by our S/N) on the \hbbc\ profile
attributable to radio loudness,  and that {\em Pop. B} RL and RQ sources can
have the same \Mbh\ and \lm\ values. Therefore, {\em it is not unreasonable to
conclude that a similar range of M and L/M is physically possible for both RQ
and RL sources.} However, this does not mean that the mass function for RL and
RQ sources is necessarily the same. A robust inference from the bootstrap
analysis reported in \S \ref{boo} is that the mass function and the
conditional probability of having certain L/M values at fixed M are likely to
be different for the two AGN classes. A correct estimation of the mass
function and of the L/M probability distribution demands a more thorough
analysis that includes selection biases. In addition, the RL/RQ dichotomy is
probably related to parameters other than M and L/M, such as BH spin and
morphological segregation. We suggest that the intrinsic mass function and L/M
distribution differences as well as jet-related effects like \oiiiopt\
enhancement in RL sources and sample selection criteria may drive the RQ/RL
separation shown by Laor (2000) and Boroson (2002).


\subsection{Transition in \hbbc\ Line Profile Shape}

Connecting empirical and physical parameters, we find that median profiles
for  \hbbc\ in  \PopA\ sources ($\log \rm L/M > 3.9$, and for   $3.5 < \log
\rm L/M < 3.9$ and 7.5$< \log \rm L/M <$8.5) show  a Lorentzian shape. In
other words, if $\log\rm L/M \ga$3.9, the shape is Lorentzian for all values
of M. Similarly, if $\log$\lm $\la$ 3.4 (bins dominated by  \PopB\ surces), the
profile is redward asymmetric and can be decomposed into two Gaussian
components (one redshifted and the other not). This suggests that the
transition from Lorentzian to double Gaussian profiles is governed by a
critical L/M value (it is interesting to note that the \hbbc\ asymmetry is also
one of the major correlate of the original E1 of Boroson \& Green 1992).
Unfortunately, the division of the bins in the L/M strip 3.4$< \log \rm L/M <
$3.9 did not make the situation clearer. We conclude that the Lorentzian --
double Gaussian profile transition occurs between 3.4$\la \log \rm L/M \la$3.9
(0.08 $\la$ L/L$_{\rm Edd}$ $\la$0.25). A more precise value would require
knowledge of source orientation effects and a better-constrained $\alpha$\
value. This provides a physical basis for  the phenomenological finding about
profile change at FHWM(\hbbc)$\approx$ 4000 \kms and the resulting Population
A-B hypothesis (Sulentic et al. 2002).


%
%

\subsubsection{c(1/4) \hbbc\ Dependence on Mass: Not Only Gravitational
Redshift}

We remark that a Double Gaussian fit to  \hbbc\ in  \PopB\
sources is a formal result: two components are required to
account for the redward asymmetry. The two components have a
physical justification if, for example, the broader (VBLR) one
can be ascribed to gas that lies closest to the continuum source
in an optically thin (to the HI ionizing continuum) region with
large covering factor $\rm f_c \approx$ 1 (as defined in Sulentic
\& Marziani 1993 and Brotherton 1996). A new result from this
investigation is that the amplitude of the redward asymmetry is
mass dependent.

The c(1/4) dependence on \Mbh\ (for $\Delta \log M\approx 1$\ we obtain a
factor $\approx$6 increase in the c(1/4) redward displacement) means that we
can try to ascribe the presence of the ``red shelf"/VBLR component  and the
redward asymmetry to gravitational and transverse redshift (e. g., Corbin
1997). If the VBLR redshift is gravitational and transverse taking the c(1/4)
value  as a conservative estimate, we obtain the following distances from the
central continuum source for the line emitting gas: if $3.5 \la \log \rm L/M
\la 3.9$, $\rm r \approx 0.005$\ pc and 0.01 pc for $\log \rm M =8$ and $\log
\rm M = 9$ respectively. In the case of the largest VBLR redshifts, the shift
in radial velocity is $\Delta \rm v_r\approx$2300 \kms\ and we find $\rm r
\approx 0.015$ pc $\approx$ 170 gravitational radii for $\log M = 9$. If we
model the VBLR gas as a shell ($\rm f_c \approx$1) with optical depth to the
Lyman continuum $\tau \la 1$, a {\tt CLOUDY} (Ferland 2000) simulation shows
luminosity of \hb\ $\log$L(\hb)$\rm_ {VBC} \approx 41.7$, where the line
luminosity is in ergs s$^{-1}$. This falls far short in explaining the VBLR
luminosity for sources in bin  3.5$ < \log \rm L/M <$ 3.9, $8.5 < \log \rm M <
9.5$. The average $\overline{\log \rm L}$(\hbbc) is 43.08. The VBLR
contributes 2/3 of the total, so the VBLR average luminosity is
$\overline{\log \rm L}$(\hb)$\rm_ {VBC} \approx$ 42.9. The difference between
the expected and observed VBLR luminosity is largely a consequence of the
small shell radius required to explain the large $\Delta v_r $ in the c(1/4).
In addition, the assumption of c(1/4) being dominated by gravitational +
transverse redshift implies that $ z_{\rm grav} \approx 3/2 \rm (FWHM/c)^2$,
which is not observed.  We conclude that, even if c(1/4) is mass dependent,
the c(1/4) shift amplitude cannot be explained by gravitational + transverse
redshift {\em alone}. Once the stringent (and perhaps unphysical) distances
from the central continuum sources set by gravitational redshift are relaxed,
we note that our VBLR model can explain \hb\ VBLR luminosity with reasonable
shell radii. While this issue requires further investigation, it is tempting
to suggest that we may be observing the long-sought infall, and that we are
observing it easily because of the low optical depth expected also along the
line of sight to  \hb. This would straightforwardly explain the ubiquitous
presence of a strong \hbbc\ redward asymmetries at low L/M.

\subsection{On the Nature of the Blue Outliers \label{nbo}}

We observe ``blue outliers" (following the simple wind model of
Zamanov et al. 2002) when the \oiiiopt\ lines and the peak of
\hb\ arise in different regions - \hb\ from a near face-on
accretion disk and [OIII] from a wind with velocity of about 1000
\kms. In terms of orientation these RQ sources  would be analogous
to some CD RL quasars or even to BL~Lac (blazars). Other
nearly-as-large L/M oriented sources can/must exist in our sample
but they are not ``blue outliers" for at least two reasons: (i)
they possess a well developed Narrow Line Region (NLR) which is
unlikely to show a significant blue-shifted component; (ii) they
are not oriented face-on. ~There are probably also lower L/M
sources oriented face-on, however their L/M ratio may not be
sufficient to power a wind or the wind outflow velocity may be
low (roughly $<$ 300 \kms). This could explain why the blue
outliers are exclussively sources with very high L/M. It is worth
noting that the only RL blue outlier, PKS 0736+01, is a flat
spectrum radio source, and one of the brightest High
Polarization, Optically Violently Variable sources. Its optical
continuum is dominated by a ``blazar" component as in BL Lac
(Malkan \& Moore 1986). Continuum properties suggest relativistic
beaming making this a true pole-on RL source consistent with the
interpretation of the ``blue outliers" proposed by Zamanov et al.
(2002).

In the most recent scenarios for joint AGN and galaxy evolution (e.g.,
Granato et al. 2001), the active nucleus and the galaxy evolve together,
with BH accreting matter and the galaxy making stars and supplying fuel for
the quasar. At some point, the wind from the accreting BH blows away the
matter surrounding it and a quasar emerges. The central accretion source
then appears as an unobscured quasar which lasts as long as there is fuel
in the accretion disk (Fabian 1999). In this scenario our ``blue outliers"
could represent the stage when the quasar has just emerged (cf. Krongold et
al. 2001 for a similar interpretation stemming from the analysis of NLSy1
host galaxies and environment). They could be in process of building of
their NLR.

\section{Conclusions }

In this paper we computed virial masses and Eddington ratios for
a sample of $\sim$300 AGN. We have shown that L/M seems to govern
the overall shape of the \hbbc, and we found an interesting
dependence on M of the \hbbc\ asymmetry. The transition between
sources showing Lorentzian and double Gaussian \hbbc\ profiles is
likely associated with a critical L/M value, as suggested by
Sulentic et al. (2002). L/M and M values in the range $8 \la$ M
$\la 9.5$ and $3 \la \log$ \lm $\la 4.2$ are found likely for
both RL and RQ AGN while others ($\log$ \lm $\ga 4.2$) may very
unlikely for RL sources. However, it is important to stress the
existence of a radio-loud blue outlier radiating at very high L/M
($\log$ \lm $\approx$ 4.8; other RL sources radiate at $\log$L/M
$\approx$4.4). This shows that there is no physical impossibility
for RL sources to be of large L/M and small M; they may be simply
less likely to be that way.


We confirm that the ``blue outliers" are  mainly NLSy1 sources
(Zamanov et al. 2002; see also Sulentic et al. 2000a). We
increase the number of known ``blue outliers" to 12. We show that
all blue outliers are accreting at 2 times higher Eddington ratio
than other sources with similar mass. The compactness of their
NLR  points toward very young ages (Zamanov et al. 2002),
suggesting that they may be ``fledgling" AGN as discussed for
NLSy1 by Sulentic et al. (2000a) and Mathur (2000). An
interesting issue is then whether different M and L/M
distributions may point toward evolutionary effects, i.e.,
whether some RQ sources may be the parent population for all
quasars (for example, NLSy1s, ``extreme \PopA'' sources, may
evolve into \PopB\ RQ or even RL depending on accretion of
angular momentum and/or host galaxy type/evolution).

In the future we will need of an orientation indicator for each individual
AGN. This is very important for a correct evaluation of \Mbh. The \civ\
line profile  shows promise as an orientation indicator at least for  \PopA\
sources (Richards et al. 2002). It may also provide us with important clues
about aspects of Broad Line Region structure and its dependence on L/M, and
ultimately, with a 3D observational space to uniquely map into a 3D
``physical" parameter space defined by M, L/M and  orientation.

\section*{Acknowledgments}
The authors acknowledge  support from the Italian Ministry of University and
Scientific and Technological Research (MURST) through grant Cofin
00$-$02$-$004. We also wish to thank Giovanna Stirpe for fruitful discussions
and a careful reading of the manuscript. This research has made use of the
NASA/IPAC Extragalactic Database (NED) which is operated  by the Jet
Propulsion Laboratory, California Institute of Technology, under contract with
the National Aeronautics and Space Administration.

\end{document}